\documentclass[aps,pre,showpacs,twocolumn]{revtex4}%
\usepackage{graphicx}
\usepackage{epsfig}
\usepackage[english]{babel}
\usepackage{amsmath}
\usepackage{color}
\usepackage{amsfonts}
\usepackage{amssymb}
\usepackage{youngtab}%
\newcommand{\beq}{\begin{equation}}
\newcommand{\eeq}{\end{equation}}
\begin{document}
\title{Reduced density matrix and entanglement entropy of permutationally invariant
quantum many-body systems}
\author{Vladislav Popkov$^{1,2}$, and Mario Salerno$^{1}$}
\affiliation{$^{1}$ Dipartimento di Fisica ``E.R. Caianiello'', Universit\`a di Salerno,
via ponte don Melillo, 84084 Fisciano (SA), Italy}
\affiliation{$^{2}$ Dipartimento di Fisica, Universit\`a di Firenze, via Sansone 1, 50019
Sesto Fiorentino (FI), Italy}

\begin{abstract}
In this paper we discuss the properties of the reduced density matrix of
quantum many body systems with permutational symmetry and present basic
quantification of the entanglement in terms of the von Neumann (VNE), Renyi
and Tsallis entropies. In particular, we show, on the specific example of the
spin $1/2$ Heisenberg model, how the RDM acquires a block diagonal form with
respect to the quantum number $k$ fixing the polarization in the subsystem
conservation of $S_{z}$ and with respect to the irreducible representations of
the $\mathbf{S_{n}}$ group. Analytical expression for the RDM elements and for
the RDM spectrum are derived for states of arbitrary permutational symmetry
and for arbitrary polarizations. The temperature dependence and scaling of the VNE
across a finite temperature phase transition is discussed and the RDM moments
and the R\'{e}nyi and Tsallis entropies calculated both for symmetric ground
states of the Heisenberg chain and for maximally mixed states.

\end{abstract}
\date{\today}

\maketitle

\section{Introduction}

Entanglement properties of interacting quantum many-body systems
\cite{Amico-Review} lies at the heart of many quantum information processes
such as measurement based quantum computation, teleportation, security of
quantum key distribution protocols, super-dense coding, etc. \cite{Nielsen}.
Being a principal resource for quantum information, one is interested to know
how much entanglement is present in a system and how much of it can be used or
created. Entanglement also provides benchmarks for success of quantum
experiments. Entanglement properties are presently investigated for several
spin chains \cite{Vidal}, \cite{Latorre}, \cite{Jin-Korepin}, \cite{Korepin},
\cite{Peschel2004}, \cite{Peschel2005}, \cite{Its},
\cite{Entanglement_Heisenberg}, \cite{PSS}, for strongly correlated fermions
\cite{Korepin}, \cite{Gu}, \cite{Anfossi} and pairing models \cite{Zanardi},
for itinerant bosons \cite{Helmerson}, etc.

The calculation of the entanglement involves the knowledge of the \textit{reduced density matrix}
(RDM) characterizing quantum systems in contact with the environment such as a
thermal bath or a larger system of which the original system is a subsystem.
In particular, the spectrum of the RDM, which by definition is real and
nonnegative with all eigenvalues summing up to one,  provides an intrinsic characterization
of a subsystem. The relative importance of a subsystem state, indeed, is directly related
 to the weight that the corresponding eigenvalue has in
the RDM spectrum. Thus, for instance, the fact that the eigenvalues
$\lambda_{i}$ of the RDM for a one dimensional quantum interacting subsystems
decay exponentially with $i$ implies that the properties of the subsystems are
determined by only a few states. This property is crucial for the success of
the density-matrix renormalization group (DMRG) method \cite{White92} in one
dimension. In two dimensions this property is lost \cite{PeschelChung} and the
DMRG method fails.

For a subsystem consisting of $n$ sites (or $n$ q-bits) the RDM is of rank
$2^{n}$ so that for large $n$ the calculation of the spectrum becomes a
problem of exponential difficulty. While the spectrum of the full RDM for
subsystems with a small number of sites (e.g. $n\leq6$) has been calculated
\cite{NienhuisCalabrese2009}, the full RDM for arbitrary $n$ is, to our
knowledge, exactly known only for the very special case of non interacting
quantum systems such as free fermions (see e.g. \cite{PeschelFreeFermionRDM})
or free bosons.

The aim of the present paper is to analytically calculate the elements of the
RDM of permutational invariant quantum systems of arbitrary size $L$, for
arbitrary permutational symmetry of the state of the system (labelled by an
integer number $0<r<L/2$) and arbitrary sizes $n$ (number of q-bits) of
the subsystem. We remark that the invariance under the permutational group
physically implies that the interactions among sites have infinite range. As
an example of such system we consider the Heisenberg model of spin $1/2$ on a full graph
consisting of $L$ sites, with fixed value of spin polarization $S_{z}=L/2-N$.
For this system we calculate the RDM and the entanglement von Neumann entropy (VNE) 
for a subsystem of arbitrary $n\geq1$
$\ $\ sites for arbitrary $L,N.$
The temperature dependence and the scaling properties of the VNE
across a finite temperature phase transition occurring in the system  are also  discussed,
and the RDM moments
and the R\'{e}nyi and Tsallis entropies calculated both for symmetric ground
states of the Heisenberg chain and for maximally mixed states.

The plan of the paper is the following. In Section
\ref{sec:Model equation and main definitions} we discuss model equations and
provide basic definitions. In Section \ref{sec:Properties of the RDM} we
consider the main properties of RDM elements and show how the symmetry
properties of the system allow to decompose the RDM into a block diagonal
form. In Sec. \ref{sec::Analytical expression of RDM elements} we present an
exact analytical expression of the RDM matrix elements for arbitrary
parameters values whose rigorous proof is provided in the thermodynamic limit
in the appendix \ref{app::RDM elements in the thermodynamic limit} and for the
case of fully symmetric states in the appendix
\ref{app::RDM elements for symmetric states}. In Sec.
\ref{sec::Spectral properties of RDM and entanglement entropy} we provide an
analytical characterization of the RDM spectrum and discuss scaling properties
and temperature dependence of the von Neumann entropy. In Sec.
\ref{sec::Moments of the reduced density matrix and Renyi and Tsallis entropies}%
, moments of the reduced density matrix are discussed and several quantities
of interest like mutual information, Renyi and Tsallis entropies, are
calculated. In the last section the main results of the paper are briefly summarized.

\label{sec:Model equation and main definitions}

\section{Model equation and basic definitions}

We consider a permutational invariant system of $L$ spins $1/2$ on a
complete graph with fixed total spin polarization $S_{z}=L/2-N$ and described
by the Hamiltonian
\begin{equation}
H=-\frac{J}{2L}\left(  \mathbf{S}^{2}-\frac{L}{2}\left(  \frac{L}{2}+1\right)
\right)  +hS_{z}\label{Hamiltonian}%
\end{equation}
Here $\mathbf{S}\equiv(S_{x},S_{y},S_{z}),\;S_{\alpha}=\frac{1}{2}\sum
_{i=1}^{L}\sigma_{i}^{\alpha}$, with $\sigma_{i}^{\alpha}$ Pauli matrices
acting on the factorized $\prod\limits_{1}^{L}\otimes C_{2}$ space. This
Hamiltonian is invariant under the action of the symmetric group
$\mathbf{S}_{L}$ \cite{Sagan} and conserves $S_{z}$, $[H,S_{z}]=0$. A complete
set of eigenstates of $H$ are states $|\Psi_{L,N,r}\rangle$ associated to
filled Young tableaux (YT) of type $\{L-r,r\}_{(N)}$ (see \cite{Mario94} for details),  where the subscript $N$ denotes the number of quanta present in the tableau and the symbol $\{L-r,r\}$
refers to a tableau of only two rows, with $L-r$ boxes (sites) in the first
row and $r$ in the second row. For the Hamiltonian in (\ref{Hamiltonian}) we
have:
\begin{align}
&  H|\Psi_{L,N,r}\rangle=E_{L,N,r}|\Psi_{L,N,r}\rangle,\nonumber\\
&  E_{L,N,r}=\frac{1}{2}\left(  \frac{Jr}{L}(L-r+1)+h(L-2N)\right)
,\label{Hamiltonian_spectrum}\\
&  S_{z}|\Psi_{L,N,r}\rangle=\left(  \frac{L}{2}-N\right)  |\Psi
_{L,N,r}\rangle\nonumber
\end{align}
where $N=0,1,...,L$ determines possible values of the spin polarization and
$r$ takes values $r=0,1,...,\max(N,L-N)$. Notice that, due to the symmetry and
antisymmetry of a YT with respect to rows and columns, respectively, the state
$|\Psi_{L,N,r}\rangle$ can exist only if $N\geq r$ (for the explicit form of
the state $|\Psi_{L,N,r}\rangle$ see Eq. (\ref{YoungTableau_example}) below).
The degeneracies of the eigenvalues $E_{L,N,r}$ are given by the dimension of
the corresponding YTs:
\begin{equation}
\deg_{L,r}=\binom{L}{r}-\binom{L}{r-1}.
\end{equation}

\noindent Consider a set of vectors $|\Psi_{u}\rangle$, $u=1,...\deg_{L,r}$,
forming an orthonormal basis in the eigenspace of $H$ with eigenvalue
$E_{L,N,r}$. We define the density matrix of the whole system as
\begin{equation}
\sigma_{L,N,r}=\frac{1}{\deg_{L,r}}\sum_{u=1}^{\deg_{L,r}}|\Psi_{u}%
\rangle\langle\Psi_{u}|. \label{thermal_density_matrix}%
\end{equation}

It can be easily shown that $\sigma_{L,N,r}$ possess the following properties:

\noindent$\mathbf{i)}$ The matrix $\sigma_{L,N,r}$ has eigenvalues
$\lambda_{1}=\lambda_{2}=...=\lambda_{\deg_{L,r}}=(\deg_{L,r})^{-1}$, with
remaining $2^{L}-\deg_{L,r}$ eigenvalues all equal to zero. This follows from
the fact that each vector $|\Psi_{u}\rangle$ is an eigenvector of
$\sigma_{L,N,r}$ with eigenvalue $\frac{1}{\deg_{L,r}}$. Since the spectrum of
$\sigma_{L,N,r}$ is real and nonnegative with all eigenvalues summing up to
$1$, the remaining $2^{L}-\deg_{L,r}$ eigenvalues must vanish.

\noindent$\mathbf{ii)}$ Matrix $\sigma_{L,N,r}$ satisfies: $(\sigma
_{L,N,r})^{2}=\frac{1}{\deg_{L,r}}\sigma_{L,N,r}$. This follows from the
definition (\ref{thermal_density_matrix}) and the orthonormality condition
$\langle\Psi_{w}|\Psi_{u}\rangle=\delta_{uw}$.

\noindent$\mathbf{iii)}$ Matrices $\sigma_{L,N,r}$ commute with each other
$[\sigma_{L,N,r} \,,\, \sigma_{L,N^{\prime},r^{\prime}}] = 0$. This follows
from orthogonality of eigenspaces of $H$ for different eigenvalues.

\noindent$\mathbf{iv)}$Introducing the operator $P_{ij}$, permuting subspaces
$i$ and $j$ of the Hilbert space $\prod\limits_{1}^{L}\otimes C_{2}$ on which
the matrix $\sigma_{L,N,r}$ acts, we have that:  $[\sigma,P_{ij}]=0$ for any $i,j$.

\noindent This last property can be proved by considering
\begin{align}
P_{ij}\sigma_{L,N,r}P_{ij}  &  =\frac{1}{\deg_{L,r}}\sum_{u=1}^{\deg_{L,r}%
}P_{ij}|\Psi_{u}\rangle\langle\Psi_{u}|P_{ij}\nonumber\\
&  =\frac{1}{\deg_{L,r}}\sum_{u=1}^{\deg_{L,r}}|\Psi_{u}^{\prime}%
\rangle\langle\Psi_{u}^{\prime}|.
\end{align}
The vectors $|\Psi_{u}^{\prime}\rangle=P_{ij}|\Psi_{u}\rangle$ form an
orthonormal basis, being $\langle\Psi_{w}^{\prime}|\Psi_{u}^{\prime}%
\rangle=\langle\Psi_{w}|P_{ij}^{T}P_{ij}|\Psi_{u}\rangle=\langle\Psi_{w}%
|\Psi_{u}\rangle=\delta_{uw}$, because $P_{ij}^{T}=P_{ij}$, and $(P_{ij}%
)^{2}=I$. Now, the sum $\sum_{u=1}^{\deg_{L,r}}|\Psi_{u}\rangle\langle\Psi
_{u}|=I_{\deg_{L,r}}$ is a unity operator in a factor space of dimension
$\deg_{L,r}$, and therefore it does not depend on the choice of the basis.
Note that vector $|\Psi_{u}^{\prime}\rangle$ belongs to the same factor space
as $|\Psi_{u}\rangle$, because permutation $P_{ij}$ only results in different
enumeration. Consequently,
\begin{equation}
P_{ij}\sigma P_{ij}=\rho\text{, \ \ \ or \ }[\sigma,P_{ij}]=0.
\label{Prho-rhoP}%
\end{equation}
The latter property implies that in Eq. (\ref{thermal_density_matrix}) the sum
over the orthogonalized set of basis vector in (\ref{thermal_density_matrix})
can be replaced by the symmetrization of the density matrix directly, namely
$\sigma_{L,N,r}=\frac{1}{L!}\sum_{P}|\Psi_{12...L}\rangle\langle\Psi
_{12...L}|$, where the sum is over all $L!$ permutations of indexes
$1,2,...L$, and $|\Psi_{12...L}\rangle$ is some unit eigenvector of $H$ with
eigenvalue $E_{L,N,r}$. In particular, it is convenient to choose
$|\Psi_{12...L}\rangle\equiv|\Psi_{L,N,r}\rangle$,
\begin{equation}
\sigma_{L,N,r}=\frac{1}{L!}\sum_{P}|\Psi_{L,N,r}\rangle\langle\Psi_{L,N,r}|.
\label{sigma_as_SumOfPermutaitons}%
\end{equation}
It is evident that such a sum is invariant with respect to permutations and
that $\sigma_{L,N,r}$ is properly normalized: $Tr\sigma_{L,N,r}=1$.

The Reduced Density Matrix (RDM) of a subsystem of $n$ sites is defined by
tracing out $L-n$ degrees of freedom from the density matrix of the whole
system:
\begin{equation}
\rho_{(n)}=Tr_{L-n}\sigma_{L,N,r}. \label{RDM}%
\end{equation}
Due to the properties (\ref{Prho-rhoP}) and (\ref{RDM}), $\rho_{(n)}$ does not
depend on the particular choice of the $n$ sites, and satisfies the property
(\ref{Prho-rhoP}) in its subspace (we omit the explicit dependence of
$\rho_{(n)}$ on $L,N,r$ for brevity of notations).

\label{sec:Properties of the RDM}

\section{RDM properties and block diagonal form}

The RDM can be calculated in the natural basis by using its definition in
terms of observables: $\langle\hat{f}\rangle=Tr(\rho_{(n)}\hat{f})$ where
$\hat{f}$ is a physical operator acting on the Hilbert space of the
$2^{n}\times2^{n}$ subsystem. The knowledge of the full set of observables
determines the RDM uniquely. Indeed, if we introduce the natural basis in the
Hilbert space of the subsystem, $\prod\limits_{k=1}^{n}\otimes C_{2}$, the
elements of the RDM in this basis are
\begin{equation}
\rho_{j_{1}j_{2}...j_{n}}^{i_{1}i_{2}...i_{n}}=\left\langle \hat{e}%
_{j_{1}j_{2}...j_{n}}^{i_{1}i_{2}...i_{n}}\right\rangle =Tr\left(  \rho
_{(n)}\hat{e}_{j_{1}j_{2}...j_{n}}^{i_{1}i_{2}...i_{n}}\right)  ,
\end{equation}
with $\hat{e}_{j_{1}j_{2}...j_{n}}^{i_{1}i_{2}...i_{n}}=\prod\limits_{k=1}%
^{n}\otimes\hat{e}_{j_{k}}^{i_{k}}$ and $\hat{e}_{j}^{i}$ a $2\times2$ matrix
with elements $\left(  \hat{e}_{j}^{i}\right)  _{kl}=\delta_{ik}\delta_{jl}$.
The matrix $\hat{e}_{j_{1}j_{2}...j_{n}}^{i_{1}i_{2}...i_{n}}$ has only one
nonzero element, equal to $1$, at the crossing of the row $2^{n-1}%
i_{1}+2^{n-2}i_{2}+...+i_{n}+1$ and the column $2^{n-1}j_{1}+2^{n-2}%
j_{2}+...+j_{n}+1$ (all indices $i,j$ take binary values $i_{k}=0,1$ and
$j_{k}=0,1$). To determine all the RDM elements one must find a complete set
of observables and compute the averages $\left\langle \hat{e}_{j_{1}%
j_{2}...j_{n}}^{i_{1}i_{2}...i_{n}}\right\rangle $. Note that a generic
property of the RDM elements, which follows directly from (\ref{Prho-rhoP}),
is that any permutation between pairs of indices $(i_{m},j_{m})$ and
$(i_{k},j_{k})$ does not change its value, e.g.
\begin{equation}
\rho_{j_{1}j_{2}...j_{n}}^{i_{1}i_{2}...i_{n}}=\rho_{j_{2}j_{1}...j_{n}%
}^{i_{2}i_{1}...i_{n}}=\rho_{j_{n}j_{1}...j_{2}}^{i_{n}i_{1}...i_{2}}%
=...=\rho_{j_{n}j_{n-1}...j_{1}}^{i_{n}i_{n-1}...i_{1}}.
\label{RDM_elements_Permute}%
\end{equation}
Another property of the RDM follows from the $S_{z}$ invariance
\begin{equation}
\rho_{j_{1}j_{2}...j_{n}}^{i_{1}i_{2}...i_{n}}=0,\text{\qquad if }i_{1}%
+i_{2}+...+i_{n}\neq j_{1}+j_{2}+...+j_{n}. \label{RDM_elements_IceRule}%
\end{equation}
Thus, for instance, the RDM for $n=2$ has only $6$ nonzero ($4$
different)\ elements , $\rho_{00}^{00},\rho_{01}^{10}=\rho_{10}^{01},$
$\rho_{10}^{10}=\rho_{01}^{01}$ and $\rho_{11}^{11}$, subject to normalization
$Tr\rho_{(2)}=\rho_{00}^{00}+$ $\rho_{10}^{10}+\rho_{01}^{01}+\rho_{11}%
^{11}=1$. It is convenient to introduce the operators
\begin{align}
&  \hat{e}_{0}^{1}=%
\begin{pmatrix}
0 & 0\\
1 & 0
\end{pmatrix}
\equiv\sigma^{-},\;\;\;\;\;\hat{e}_{1}^{0}=%
\begin{pmatrix}
0 & 1\\
0 & 0
\end{pmatrix}
\equiv\sigma^{+},\nonumber\\
&  \hat{e}_{0}^{0}=%
\begin{pmatrix}
1 & 0\\
0 & 0
\end{pmatrix}
\equiv\hat{p}\;,\;\;\;\;\;\hat{e}_{1}^{1}=%
\begin{pmatrix}
0 & 0\\
0 & 1
\end{pmatrix}
\equiv\hat{h}\;.\nonumber
\end{align}
If we represent a site spin up with the vector $\binom{1}{0}$ and a site spin
down with the vector $\binom{0}{1}$ then $\hat{p}_{k}$ and $\hat{h}_{k}$ are
spin up and spin down number operators on site $k$, while $\sigma^{-}%
,\sigma^{+}$ represent spin lowering and rising operators, respectively. Thus,
for instance, the observable $\langle\hat{p}_{1}\hat{p}_{2}\hat{h}_{3}\hat
{h}_{4}...\hat{h}_{n}\rangle=$ $\rho_{0011...1}^{0011...1}$ gives the
probability to find spins down at sites $3,4,...n$, and spins up at sites
$1,2$, while the observable $\langle\sigma_{1}^{+}\sigma_{2}^{+}\sigma_{3}%
^{-}\sigma_{4}^{-}\hat{h}_{5}...\hat{h}_{n}\rangle=\rho_{11001...1}%
^{00111...1}$ gives the probability to find spins down at sites $5,6,...n$,
spin lowering at sites $3,4$ and spin rising at sites $1,2$. Note that the
latter operator conserves the total spin polarization since the number of
lowering and rising operators is the same. Also note that the correlation
functions with a non conserved polarization vanish, e.g.
\begin{equation}
\langle\sigma_{1}^{+}\sigma_{2}^{+}\sigma_{3}^{-}\hat{h}_{4}\hat{h}_{5}%
...\hat{h}_{n}\rangle=\rho_{11011...1}^{00111...1}=0,
\end{equation}
in accordance with (\ref{RDM_elements_IceRule}).

%\label{sec:Block diagonal form of the RDM}

%\section{of the RDM}

\begin{figure}[h]
\centerline{\scalebox{.9}{\includegraphics{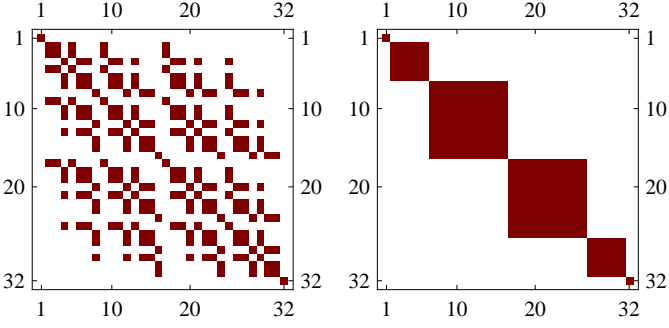}}
}\caption{Left panel. RDM $\rho_{(5)}$ in the natural basis. Parameters are
$L=18,N=8,r=6$. Black boxes denote non zero elements. Right panel. The same
matrix of the left panel after the following chain of permutations of rows and
columns have been applied to it: $R_{8,14}R_{14,18} R_{8,21}R_{3,29}%
R_{2,28}R_{6,16}R_{4,9}$. ($R_{i,j}$ denote the operator which exchange first
columns $i$ and $j$ and then rows $i$ and $j$). Black boxes denote non zero
elements. Block diagonal structure associated with values of $k=0,1,...,5$ is
evident. The single element present in blocks $k=0, 5,$ are $1/34$ and
$1/153$, respectively. Elements values inside other k-blocks are given in Fig.
\ref{Fig_n6_threeblocks}. }%
\label{Fig_rho6_sparse}%
\end{figure}

One can take advantage of the $S_{z}$ invariance (e.g. Eq.
\ref{RDM_elements_IceRule}) to block diagonalize the RDM into independent
blocks $B_{k}$ of fixed polarization $k=i_{1}+i_{2}+...+i_{n}=j_{1}%
+j_{2}+...+j_{n}$ (here $k=0,...,n$ gives the number of spin up present in the
subsystem). In Fig. \ref{Fig_rho6_sparse} the blocks $B_{k}$ appearing in the
RDM $\rho_{(5)}$ have been shown for the case $n=5,L=18,N=8,r=6$. We remark that
the $n+1$ diagonal blocks correspond to the values $s_{z}=(n - 2 k)/2$,
$k=0,1,..., n$ the subsystem polarization can assume, being the block
decomposition a direct consequence of the $S_{z}$ symmetry. The dimension of
the block $B_{k}$ coincides therefore with the number of possible
configurations that $k$ spin up can assume on $n$ sites, e.g. $\dim
B_{k}=\binom{n}{k}$. One can check that the sum of the dimensions of all
blocks gives the full RDM dimension, i.e. $\sum_{k}\dim B_{k}=2^{n}$. Notice
that the block diagonal form in the natural basis is achieved only after a
number of permutations of rows and columns of the RDM have been performed. We
also remark that the fact that the middle block $B_{3}$ in Fig.
\ref{Fig_rho6_sparse} has all vanishing anti-diagonal elements is purely
accidental (see also remark at the end of Sec.\ref{sec:Properties of the RDM}).

\begin{figure}[h]
\centerline{\scalebox{.9}{\includegraphics{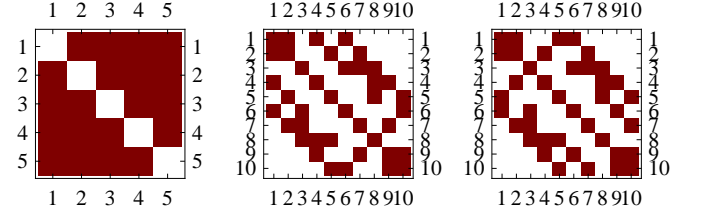}}}\caption{Blocks
$B_{k}$ for $k=1, 4$ (left panel), $k=2$ (central panel) and $k=3$ (right
panel) corresponding to the block diagonal form of $\rho(5)$ in
Fig.\ref{Fig_rho6_sparse}. In the left panel white and black boxes denote
elements $5/306$ and $1/2448$ for $k=1$ and elements $2/51$ and $1/1020$ for
case $k=4$. In the center panel white boxes denote the element $1/1020$ while
black boxes, except for the ones along the diagonal which are all equal to
$2/51$, denote the element $-1/1190$. In the right panel white boxes denote
the element $1/1360$ while black boxes denote the element $-3/4760$, except
the ones along the diagonal which are equal to $1/34$. The matrices
$\sigma_{n,k,s}$ involved in the decomposition in (\ref{B3_decomposition})
have the same shapes as blocks $B_{k}$ in the left panel for $\sigma_{511} =
\sigma_{541}$, with white and black boxes denoting elements $1/5$ and $-1/20$,
respectively. The shape of matrix $\sigma_{521}$ (resp. $\sigma_{522}$) is the
same as in the central panel, with white boxes denoting $1/60$ (resp. $-1/30$
) and black ones $- 1/15$ (resp. $1/30$ ), except the ones along the diagonal
which are $1/10$, and shape of $\sigma_{531}$ (resp. $\sigma_{532}$) is the
shape is as in right panel with white box denoting $1/60$ (resp. $-1/30$) and
black ones $-1/15$ (resp. $1/30$ ) except for the ones along the diagonal
which are given by $1/10$. The $5\times5 $matrices $\sigma_{510}=\sigma_{540}%
$, and $10 \times10$ matrices $\sigma_{520}=\sigma_{530}$ have all elements
equal to $1/5$ and $1/10$, respectively, while $\sigma_{500}=\sigma_{550}=1$.
}%
\label{Fig_n6_threeblocks}%
\end{figure}

Blocks $B_{k}$ consist of elements $e_{j_{1}}^{i_{1}}\otimes e_{j_{2}}^{i_{2}%
}\otimes e_{j_{3}}^{i_{3}}\otimes...\otimes e_{j_{n}}^{i_{n}}$ \ of the
original matrix, with $\sum\limits_{1}^{n}i_{p}=\sum\limits_{1}^{n}j_{p}=k$
and $i_{p}=0,1$, $j_{p}=0,1$. In its turn, all elements $e_{j_{1}}^{i_{1}%
}\otimes e_{j_{2}}^{i_{2}}\otimes e_{j_{3}}^{i_{3}}\otimes...\otimes e_{j_{n}%
}^{i_{n}}$ of the block $B_{k}$ can be further block diagonalized
%according
%to the irreducible representations of the permutation group $\mathbf{S_{n}}$
%of the subsystem
(see below). In the natural basis, this diagonalization is done according to
the number of pairs of type $(e_{1}^{0}\otimes e_{0}^{1})$ present in the
elements. In the following we denote by $G_{Z}$ the part of the block
associated to elements with $Z$ pairs $(e_{1}^{0}\otimes e_{0}^{1})$ in it.
The sub-block $G_{0}$ of the block $B_{k}$ is formed by the elements
containing $k$ terms $e_{1}^{1}$ and ($n-k$) terms $e_{0}^{0}$ in the product,
i.e. $e_{1}^{1}\otimes...e_{1}^{1}\otimes e_{0}^{0}\otimes...\otimes e_{0}%
^{0}$ and all permutations. All such elements lie on the diagonal, and vice
versa, each diagonal element of $B_{k}$ belongs to $G_{0}$. Consequently, the
sub-block $G_{0}$ consists of $\binom{n}{k}$ elements. The number of elements,
$\deg G_{1}(k)$, in the sub-block $G_{1}$ is equal to the number of elements
of the type $e_{1}^{0}\otimes e_{0}^{1}\otimes e_{i_{1}}^{i_{1}}\otimes
e_{i_{2}}^{i_{2}}\otimes...\otimes e_{i_{n-2}}^{i_{n-2}}$, such that
$1+0+i_{1}+i_{2}+...+i_{n-2}=k$. 
Using elementary combinatorics we obtain:
\begin{equation}
\deg G_{1}(k)=\binom{2}{1}\binom{n}{2}\binom{n-2}{k-1}.
\end{equation}
Analogous calculations for arbitrary sub-block $G_{Z}$ yields
\begin{equation}
\deg G_{Z}(k)=\binom{2Z}{Z}\binom{n}{2Z}\binom{n-2Z}{k-Z}.
\label{deg_Gz(k)}%
\end{equation}
From the restriction $\sum_{1}^{n}i_{p}=\sum_{1}^{n}j_{p}=k$ we
deduce that the block $B_{k}$ contains non-empty parts $G_{0},G_{1}%
,...,G_{\min(k,n-k)}$, leading to the following decomposition:
\begin{equation}
B_{k}=\bigcup\limits_{Z=0}^{\min(k,n-k)}G_{Z}. \label{Bk_decomposition}%
\end{equation}
Indeed, the normalization condition following from (\ref{Bk_decomposition}),
gives
\begin{equation}
\sum_{Z=0}^{\min(k,n-k)}\deg G_{Z}(k)=\binom{n}{k}^{2}.
\end{equation}
It is important to note that \textit{all elements of} $G_{Z}$ \textit{are
equal}. This is a direct consequence of the property in Eq.
(\ref{RDM_elements_Permute}). A graphical representation of the $B_{k}$ block
for a particular choice of $L,N,r$, is given in Fig.\ref{Fig_n6_threeblocks}.

It is instructive to discuss the structure of blocks $B_{k}$ in terms of the
matrices $\sigma$ in in (\ref{sigma_as_SumOfPermutaitons}) since this
structure is directly connected with the block diagonalization of the RDM with
respect the the irreps of \textbf{$S_{n}$}. 
We have, indeed, that each block
$B_{k}$ can be decomposed in the form
\begin{equation}
B_{k}=\sum\limits_{s=0}^{\min(k,n-k)}\alpha_{n,k,s}^{L,N,r}\sigma_{n,k,s}
\label{Bk_decomposition_sigma}%
\end{equation}
where $\sigma_{n,k,s}$ are associated to filled YTs of type $\{n-s,s\}_{(k)}$
and $\alpha_{n,k,s}^{L,N,r}$ are coefficients related to the corresponding
eigenvalues $\lambda_{n,k,s}^{L,N,r}$ of the RDM by
\begin{equation}
\alpha_{n,k,s}^{L,N,r}= \lambda_{n,k,s}^{L,N,r} \deg_{n,s} .
\end{equation}

Notice that the matrices $\sigma_{n,k,s}$ in the natural basis have dimension
$2^{n}$ and coincide with the ones given in (\ref{thermal_density_matrix}) .
In the proper basis (e.g. that of the irrep of \textbf{$S_{n}$}) they have
dimension $\deg_{n,s} \times\deg_{n,s}$ and contribute to $B_{k}$ with a
sub-block of dimension $\deg_{n,s}$ corresponding to the filled YT of type
$\{n-s,s\}_{(k)}$. In performing this reduction one actually achieves the
diagonalization of the block $B_{k}$, as it is evident from
(\ref{Bk_decomposition_sigma}) (recall that $\sigma_{n,k,s}$ have eigenvalues
$1/\deg_{n,s}$). A first reduction of the matrices is achieved by accounting
for the $S_{z}$ symmetry discussed before, this leading to matrices
$\sigma_{n,k,s}$ of size $\binom{n}{k} \times\binom{n}{k}$. In Fig.
\ref{Fig_n6_threeblocks} are also given the matrices
$\sigma_{n,k,s}$ appearing in the decomposition of blocks $B_k$
for the specific example considered in Fig. \ref{Fig_rho6_sparse} and for
$k=0,...5, \;s=0,1,2$.
One can check that these matrices satisfy all
properties of matrix $\sigma$ given above and in particular, the number of
their nonzero eigenvalues (all equal to $1/\deg_{n,s}$) coincides with the
dimension $\deg_{n,s}$ of the YT to which they are associated. This implies
that they can be further reduced from $\binom{n}{k}\times\binom{n}{k} $ to
$\deg_{n,s}\times\deg_{n,s}$ size by eliminating the spurious $\binom{n}{k}-
\deg_{n,s}$ zero eigenvalues (these eigenvalues arise because in the natural
basis the dimension of the representation is larger than the one of the
\textbf{$S_{n}$} irreps). This is achieved by using the singular valued
decomposition of the matrix $\sigma$ to write it in the form: $\sigma= U W
V^{T}$, where $W$ a diagonal matrix whose elements are the singular values and
$U$ and $V$ are orthogonal matrices: $U^{T} U = V^{T} V =1$, with superscript
$T$ denoting the transpose (this decomposition can be obtained very
efficiently numerically \cite{numrecepies}).

The reduction to the sub-blocks of $B_{k}$ in the proper \textbf{$S_{n}$}
representation is then achieved as: $\sigma= u w v^{T}$ where $u$ and $v$ are
rectangular matrices of dimension $\binom{n}{k} \times\deg_{n,s}$ obtained
from $U$ and $V$ by omitting the columns corresponding to the zero eigenvalues
and the matrix w is a $\deg_{n,s} \times\deg_{n,s}$ diagonal matrix with the
nonzero eigenvalues along the diagonal (in our case, since the nonzero
eigenvalues of $\sigma$ are all equal to 1, $w$ reduces to an unit matrix).
The matrix $w$ then provides the representation of $\sigma_{n,k,s}$ in the
proper \textbf{$S_{n}$} space leading to the full diagonalization of the block
$B_{k}$.

Thus, for example, the block diagonal form of the RDM in the right panels of
Fig. \ref{Fig_rho6_sparse} (see also Fig.\ref{Fig_n6_threeblocks}) is
expressed in terms of matrices $\sigma$ as
\begin{align}
\label{B3_decomposition}B_{1}  &  = \lambda_{510} \sigma_{510} \oplus4
\lambda_{511} = \lambda_{540} \sigma_{540} \oplus4\; \lambda_{541}
\sigma_{541} = B_{4},\nonumber\\
B_{2}  &  = \lambda_{520}\,\sigma_{520} \oplus4\; \lambda_{521}\,\sigma_{521}
\oplus5\; \times\lambda_{522} \,\sigma_{522},\\
B_{3}  &  = \lambda_{530}\,\sigma_{530} \oplus4\; \lambda_{531}\,\sigma_{531}
\oplus5\; \times\lambda_{532} \,\sigma_{532},\nonumber\\
\end{align}
with blocks $B_{0} = \lambda_{500} =\frac{1}{34}$, $B_{5} = \lambda_{550} =
\frac{1}{153}$ and eigenvalues $\lambda_{510}=11/612$, $\lambda_{520}%
=76/1785$, $\lambda_{530}=19/595$, $\lambda_{540}=11/255$, of degeneracy $1$,
eigenvalues $\lambda_{511}=13/816$, $\lambda_{521}=299/7140$, $\lambda
_{531}=299/9520$, $\lambda_{541}=13/340$, of degeneracy $4$, and eigenvalues
$\lambda_{522}=13/357$, $\lambda_{532}=13/476$, of degeneracy $5$ (having
adopted the short notation $\lambda_{nks} \equiv\lambda^{L,N,r}_{n,k,s}$, the
chosen parameters $L=18, N=8, r=6$ are understood).

\section{Analytical expression of RDM elements}

\label{sec::Analytical expression of RDM elements}
The main analytical property of the RDM is summarized in the following
statement:

\noindent\textit{Elements $g_{Z}$ of a sub-block $G_{Z}$ of a block $B_{k}$ of
the RDM (\ref{RDM}), for arbitrary $L,N,r,n$, are given by:}
\begin{equation}
g_{Z}=\frac{\binom{L-n}{N-k}}{\binom{L}{N} \binom{N}{Z} \binom{L-N}{Z}}
\sum_{m=0}^{Z}\left(-1\right)^{m}\binom{N-r}{Z-m}\binom{L-N-r}{Z-m}
\binom{r}{m}.
\label{g_z}
\end{equation}
This expression has been derived by extrapolating exact results obtained for
finite size calculations using symbolic programs and its correctness has been
checked by comparing with brute force numerical calculation of the RDM up to
large sizes. Notice that Eq. (\ref{g_z}) completely defines all elements of
the RDM in the natural basis. In practice, to find the element $P,Q$ of the
RDM $\left(  \rho_{(n)}\right)  _{PQ}$ in the natural basis one must take the
binary representation of numbers $P-1$ and $Q-1$ (which provide the sets of
integers $\{i_{p}\}$ and $\{j_{p}\}$, respectively), find the corresponding
number $Z$ and use (\ref{g_z}).

A proof of the statement for
arbitrary $L,N,n$ is given in Appendix A for the specific case $r=0$ corresponding to
fully symmetric states. A proof of Eq. (\ref{g_z}) which is valid in
the thermodynamical limit $L\rightarrow\infty$ is provided in Appendix B. In this respect, we remark that in the limit $L\rightarrow\infty$ Eq. (\ref{g_z}) simplifies to
\begin{equation}
g_{Z}=p^{n-k}(1-p)^{k}\eta^{Z}, \label{Gz_elements_thermodynamic_limit}%
\end{equation}
where we denote with $p=\frac{N}{L}$, $\mu=\frac{r}{L}$ and
\begin{equation}
\eta=\frac{(p-\mu)(1-p-\mu)}{p(1-p)}. \label{eta}%
\end{equation}
For a proof of  Eq (\ref{Gz_elements_thermodynamic_limit}) see
Appendix \ref{app::RDM elements in the thermodynamic limit}.

\section{Spectral properties of RDM and entanglement entropy}

\label{sec::Spectral properties of RDM and entanglement entropy}

The existence of two representations for the block $B_{k}$ of the RDM, one
in terms of matrices $G_{Z}$ given in Eq. (\ref{Bk_decomposition}), the other
involving matrices $\sigma_{n,k,s}$ and given in Eq.
(\ref{Bk_decomposition_sigma}), have been shown in
Sec.\ref{sec:Properties of the RDM}. These representations, together with the invariance
of $G_{Z}$ and $\sigma_{n,k,s}$ with respect to permutations, imply the
existence of linear relations of the form
\begin{equation}
\hat{G}_{Z}(k)=\sum_{s=0}^{\min(k,n-k)}\beta_{Z}(k,s)\sigma_{n,k,s}%
\end{equation}
where $\beta_{Z}(k,s)$ are constants and $\hat{G}_{Z}$ denotes the matrix
formed by all elements of $G_{Z}$. Since $\sigma_{n,k,s}$ commute for
different $s$ (see the property $(\mathbf{iii)}$ of matrices $\sigma$ in Sec.
\ref{sec:Model equation and main definitions}), we have that also $\hat{G}%
_{Z}$ commute
\begin{equation}
\lbrack\hat{G}_{Z}(k),\hat{G}_{Z^{\prime}}(k)]=0.\label{Gz_commutation}%
\end{equation}
This also implies that all RDM eigenvalues $\lambda_{n,k,s}^{L,N,r}$ must be
linear combinations of elements $g_{Z}$ of matrices $G_{Z}$. One can show,
indeed, that the general expression of the RDM eigenvalues is
\begin{equation}
\lambda_{n,k,s}^{L,N,r}=\sum_{Z=0}^{\min(k,n-k)}\alpha_{Z}^{(s)}%
(n,k)g_{Z}, \label{eigenvalue_basis_coefficients}%
\end{equation}
with coefficients $\alpha_{Z}^{(s)}(n,k)$ given by
\begin{equation}
\alpha_{Z}^{(s)}(n,k)=(-1)^{Z}\sum_{i=0}^{k-s}(-1)^{i}\binom{k-s}{i}%
\binom{n-k+s}{i}\binom{s}{Z-i},\label{alpha_Z_general}%
\end{equation}
where $Z,s=0,1,...,k$. From Eq. (\ref{alpha_Z_general}) one can see that
$\alpha_{Z}^{(s)}(n,k)$ are \textit{integer coefficients} which, due to the
property (\ref{Gz_commutation}), do not depend on the characteristics of the
original state $L,N$ and $r$. Thus the dependence of the RDM eigenvalues on
these parameters enters only through the elements $g_{Z}$ (\ref{g_z}).
Moreover, one can shown that they satisfy the following relations
\begin{align}
&  \alpha_{0}^{(s)}(n,k)=1,\label{alpha_0_property}\\
&  \sum_{Z=0}^{\min(k,n-k)}\alpha_{Z}^{(0)}(n,k)=\binom{n}{k},\label{b2}\\
&  \sum_{Z=0}^{\min(k,n-k)}\alpha_{Z}^{(s)}(n,k)=0\;\quad\text{for
}s>0,\label{b1}\\
&  \sum_{Z=0}^{\min(k,n-k)}\alpha_{Z}^{(s)}(n,k)\binom{Z}{p}%
=0,\;p=0,1,...s-1,\label{alpha_Z_property(d)}\\
&  \sum_{Z=0}^{\min(k,n-k)}\alpha_{Z}^{(s)}(n,k)\binom{Z}{k}=\alpha_{k}%
^{(s)}(n,k)=(-1)^{s}\binom{n-k-s}{k-s}.\label{alpha_Z_property(e)}%
\end{align}
a proof of which can be found for special cases in \cite{acta}.

From Eqs. (\ref{g_z}), (\ref{Gz_elements_thermodynamic_limit}), (\ref{eigenvalue_basis_coefficients}), (\ref{alpha_Z_general}), the explicit analytical form of
the \textit{complete spectrum of the RDM} is obtained.

The knowledge of the RDM spectrum allows to investigate the bipartite entanglement, e.g. the
entanglement of a subsystem of size $n$ with respect to the rest of the system
(see \cite{Amico-Review} for a review). This is done in terms of the
entanglement entropy which for pure states at zero temperature coincides with
the non Neumann entropy
\begin{equation}
S_{(n)}=-tr(\rho_{(n)}\log_{2}\rho_{(n)})=-\sum\lambda_{k}\log_{2}\lambda
_{k},\label{von Neumann_entropy}%
\end{equation}
where $\lambda_{k}$ the eigenvalues of the RDM $\rho_{(n)}$, obtained from the
density matrix $\rho$ of the whole system as $\rho_{(n)}=tr_{(L-n)}\rho$. For
the infinite range ferromagnetic Heisenberg model at zero temperature the
density matrix of the whole system is a projector on the symmetric ground
state $\rho=|\Psi(L,N)\rangle\langle\Psi(L,N)|$ considered in
\cite{Entanglement_Heisenberg} where it was shown that $\lambda_{k}=(_{k}%
^{n})(_{N-k}^{L-n})/(_{N}^{L})$, where $k=0,1,...\min(n,N)$. In the limit of
large $n$ the VNE becomes
\begin{equation}
S_{(n)}\approx\frac{1}{2}\log_{2}(2\pi epq)+\frac{1}{2}\log_{2}\frac
{n(L-n)}{L}.\label{entropy_L}%
\end{equation}
One can show that a zero temperature (e.g. $\mu=r/L=0$)
\cite{Entanglement_Heisenberg} the spectrum of the reduced density matrix is
described by a binomial distribution $\lambda_{k}=p^{k}q^{n-k}\binom{n}{k}$
which converges to a Gaussian for large $n$
\begin{equation}
\lambda_{k}\approx\frac{1}{\sqrt{2\pi\sigma^{2}}}\exp\left(  -\frac
{n^{2}(p-\frac{k}{n})^{2}}{2\sigma^{2}}\right)  ,\label{GaussianApproxT=0}%
\end{equation}
where $\sigma^{2}=np(1-p)\gg1$.
\begin{figure}[h]
\centerline{\scalebox{.8}{\includegraphics{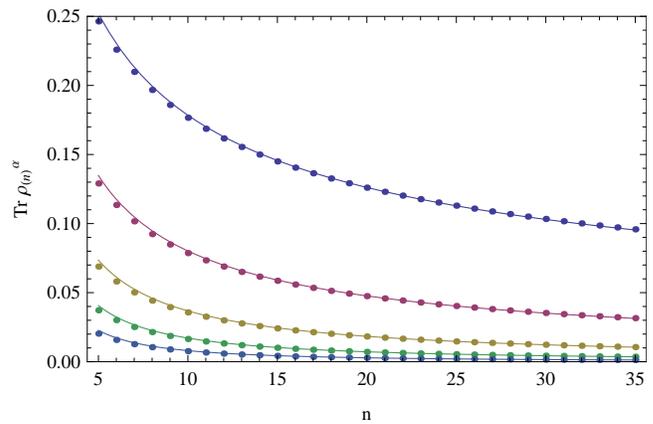}}
}\caption{Traces of the powers $\alpha=2, 2.5, 3, 3.5, 4$ (from top to bottom,
respectively) of the RDM versus the block size $n$ for symmetric states $r=0$,
as obtained from the analytical expression (\ref{tra_symm}) for $p=0.5$
(continuous curves) and from exact expressions of RDM eigenvalues
(\ref{eigenvalue_basis_coefficients}), (\ref{alpha_Z_general}) (full dots) for
$L=2\,* 10^{3}$. }%
\label{fig_power}%
\end{figure}
For finite temperature one introduces the thermal VNE for a block of size $n$
as
\begin{align}
&  S_{(n)}(\beta)=-tr(\tilde\rho_{(n)}\left(  \beta\right)  \log_{2}
\tilde\rho_{(n)}\left(  \beta\right)  ),\label{S_n_finite_size}\\
&  \tilde\rho_{(n)}\left(  \beta\right)  =\frac{1}{\mathcal{Z}}\sum_{r}
d_{r}e^{-\beta E_{r}}\langle\rho_{(n)}(r)\rangle, \label{robeta}%
\end{align}
where $\tilde\rho_{(n)}(\beta)$ is the thermal reduced density matrix,
$\mathcal{Z}$ is the partition function, and $\langle.\rangle$ denotes the
equal weight (thermic) average over all orthogonal degenerate states,
corresponding to a given permutational symmetry. Note that $\langle\rho
_{(n)}(r)\rangle$ commutes with any permutation operator and does not depend
on the choice of sites in the block but only on its size $n$. Also note that
the matrices $\langle\rho_{(n)}(r)\rangle$ commute for different $r$
\begin{equation}
\left[  \langle\rho_{(n)}(r)\rangle, \langle\rho_{(n)}(r^{\prime}%
)\rangle\right]  =0, \label{commutator}%
\end{equation}
so that the diagonalization of $\tilde\rho_{(n)}(\beta)$ is reduced to the
diagonalization of $\langle\rho_{(n)}(r)\rangle$ for arbitrary $r$. From Eq.
(\ref{robeta}) the computation of the temperature-dependent von Neumann
entropy is easily made with the help of the general expression of the
eigenvalues of the RDM in Eq.s (\ref{eigenvalue_basis_coefficients}%
),(\ref{alpha_Z_general}) for states of arbitrary permutational symmetry.
While $p=N/L$ is the system polarization, the relation between the
temperature $T$ and the parameter $\mu=r/L$ is fixed by the condition of the
minimum of the free energy of the whole system defined by the spectrum
(\ref{Hamiltonian_spectrum}) and its degeneracy. It has the form (see
\cite{EntanglementThermic}, \cite{Entanglement_Theorems})
\begin{equation}
\frac{J}{2T}=\frac{1}{(1-2\mu)}\ln\left(  \frac{1-\mu}{\mu}\right)  .
\end{equation}

The scaling of the thermal VNE across a phase transition, which occurs in the system
with infinite range interactions  at finite temperature
$T_{c} \neq0$ \cite{Mario94}, has been considered in
Ref. \cite{EntanglementThermic}, \cite{Entanglement_Theorems}. In this case it
was shown that the VNE of a block od size $n$ scales as
\begin{equation}
S_{(n)}=\left\{
\begin{array}
[c]{cc}%
\frac{1}{2}\log n+\frac{1}{2}\log2\pi e p q & \text{for }T=0\\
nH(\mu)+\frac{1}{2}\log n+C(p,\mu) & \text{ \ \ \ \ \ \ \ for }0<T<T_{c}\\
nH(\min(p,1-p)) & \text{ for }T\geq T_{c}%
\end{array}
\right.  \label{VNE(T)}%
\end{equation}
where $H(a)=-a\log a-(1-a)\log(1-a)$, $q=1-p$ and $C(p,\mu)$ does not depend
on $n$.
\begin{figure}[h]
\centerline{\scalebox{.8}{\includegraphics{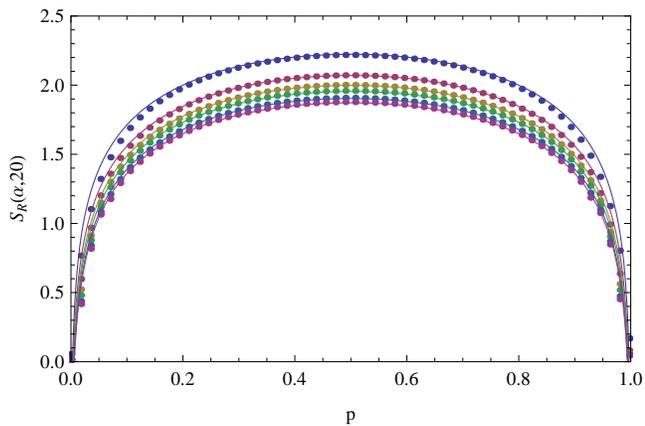}}}\caption{R\'enyi
entropies for symmetric ground states of the Heisenberg chain versus
polarization $p=N/L$ for the cases $\alpha=1,2,3,4,6,8$ (curves from top to
bottom, respectively) and for $n=20$. Continuous curves refer to the
analytical expression in (\ref{renyi-heis}) while the full dots are obtained
from exact calculations using RDM eigenvalues in
(\ref{eigenvalue_basis_coefficients}) (\ref{alpha_Z_general}) for $L=2\,*
10^{3}$ and same polarization.}%
\label{Fig_renyi}%
\end{figure}

Another quantity of interest strictly related to the entanglement entropy is
the \textit{mutual information}, $I_{AB}$, which measures the work necessary
to erase all correlations in the bipartite system \cite{Amico-Review}:
\begin{equation}
I_{AB}=S(A)+S(B)-S(AB),
\end{equation}
where $S(X)$ is the VNE of the subsystem $X$. At nonzero temperature we find
for a subsystem of size $n$ of a system with size $L$, using (\ref{VNE(T)}):
\begin{equation}
I_{AB}(n,L,T)=\frac{1}{2}\log(n(L-n))+Const
\end{equation}
for all $T<T_{c}$ and $I_{AB}(T)=0$ for $T\geq T_{c}$.

\section{Moments of the reduced density matrix and R\'enyi and Tsallis
entropies}
\label{sec::Moments of the reduced density matrix and Renyi and Tsallis entropies}%

Besides the entanglement entropy, the R\'{e}nyi, $R(\alpha)$, \cite{renyi} and
Tsallis, ($T(\alpha)$, \cite{tsallis} entropies, defined as
\begin{equation}
S_{R}(\alpha,n)=\frac{\log Tr(\rho_{(n)}^{\alpha})}{1-\alpha},\;\;S_{T}%
(\alpha,n)=\frac{Tr(\rho_{(n)}^{\alpha})-1}{1-\alpha},\label{gen_entropy}%
\end{equation}
with $\alpha$ a positive real number, are also commonly used as a measure of
entanglement. Notice that both expressions reduce the VNE in the limit
$\alpha\rightarrow1$. The knowledge of these generalized entropies requires
the computation of $Tr(\rho_{(n)}^{\alpha}$ which, except special cases (see
below) it is a very difficult task. For $\alpha$ positive integers, however,
the moments $Tr(\rho_{(n)}^{\alpha}$ can be computed using a quantum field
theory (QFT) procedure which is known as the \textit{replica method}
\cite{replica} (reminiscent of the "replica trick" of disordered systems).
\begin{figure}[h]
\centerline{
\scalebox{.8}{\includegraphics{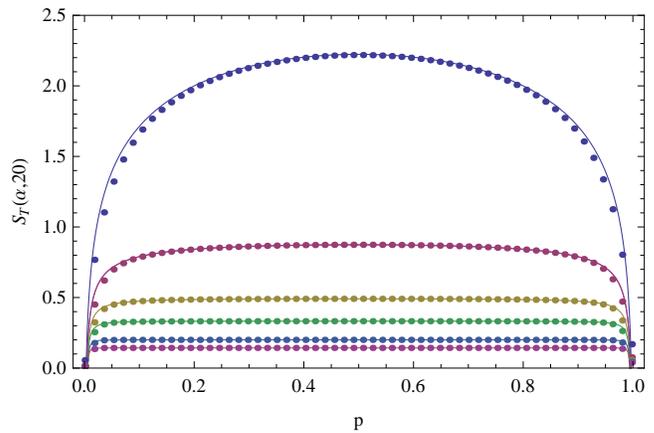}}}\caption{Same as in Fig.
\ref{Fig_renyi} but for Tsallis entropy. Continuous curves refer to the
analytical expression in (\ref{tsallis-heis}) while the full dots are obtained
from exact calculations using RDM eigenvalues.}%
\label{Fig_tsallis}%
\end{figure}
In this case the entanglement entropy is obtained through an analytical
continuation of $Tr(\rho_{(n)}^{\alpha})$ from positive integers to real
$\alpha$ values, using
\begin{equation}
S_{(}n)=-lim_{\alpha\rightarrow1}\frac{\partial}{\partial\alpha}Tr[{\rho
_{(n)}}^{\alpha}].\label{QFT}%
\end{equation}

In the case of 1+1 conformal field theories critical models at zero
temperature (for ground state) the displays universal properties, namely
\begin{equation}
Tr(\rho_{(n)}^{\alpha})=C_{\alpha}\left(  \frac{L}{\pi}\sin\frac{\pi n}%
{L}\right)  ^{-c\frac{\alpha-1/\alpha}{6}}%
\end{equation}
where $c$ is the central charge of the underlying conformal field theory.
Similarly, for the quantum XY chain with periodic boundary
conditions at zero temperature it has been shown that the RDM is independent
on the block size n and the moments can be expressed in the form
\cite{franchini}
\begin{equation}
(\frac12 Tr(\rho_{(n)}^{\alpha}))^{1/2}= \frac{\prod_{m=1}^{\infty} (1+ e^{-2
m \epsilon\alpha})}{\prod_{m=1}^{\infty} (1+ e^{-2 m \epsilon})^{\alpha}}%
\end{equation}
where $\epsilon$ depends on the anisotropy and transverse field parameters.
Except these and few other cases, analytical properties of RDM for interacting
systems are largely unexplored. The characterization of the RDM spectrum given
for permutational invariant systems allows to provide another exact result for
the RDM moments which is not accessible by QFT methods (our model is not
conformal invariant).

In particular, the ground states of the ferromagnetic Heisenberg chain being
characterized by the YT with $r=0$, are fully symmetric states with respect to
the permutations (see appendix). For these states then one can obtain the
analytical expression of $Tr(\rho^{\alpha})$ straightforwardly, using the
Gaussian distribution of the symmetric RDM eigenvalues derived in
(\ref{GaussianApproxT=0}). The approximation $\sum_{k}...\approx n\int...dx$,
indeed, readily provides
\begin{align}
Tr(\rho^{\alpha})  &  \approx n\int_{-infty}^{infty}\frac{1}{(2\pi\sigma
^{2})^{\alpha/2}}\exp\left(  -\frac{n^{2}(p-x)^{2}}{2\sigma^{2}}\alpha\right)
dx =\nonumber\\
&  = \frac{1}{\sqrt{\alpha}}\left(  2\pi pq\right)  ^{\frac{1-\alpha}{2}%
}n^{\frac{1-\alpha}{2}} \label{tra_symm}%
\end{align}
In Fig. \ref{fig_power} we compare the behavior of the traces of the RDM
powers with the block size $n$, as obtained from Eq. (\ref{tra_symm}) and from
exact expressions of RDM eigenvalues. We see that the agreement is very good
this confirming the correctness of our analytical derivation.

In the case the original global state has the form of a maximally mixed state,
i.e. is the sum of equally weighted projectors on symmetric states
$|\Psi_{L,N}\rangle$ of the form $\rho=\frac{1}{L+1}\sum_{N=0}^{L}|\Psi
_{L,N}\rangle\langle\Psi_{L,N}|$, the reduced density matrix has one
eigenvalue only $\lambda_{k}=\frac{1}{n+1}$, which is degenerate $n+1$ times.
In this case, then
\begin{equation}
Tr(\rho_{(n)}^{\alpha})=(n+1)^{1-\alpha}. \label{tra_mix}%
\end{equation}
Notice that the entanglement entropy at $T=0$ in (\ref{VNE(T)}) follows from
(\ref{tra_symm}) using the expression (\ref{QFT}) of the QFT replica method.
Another quantity directly related to the RDM moments is the \textit{effective
dimension} defined as $d_{eff}=\frac{1}{Tr(\rho_{(n)}^{2})}$. Summarizing the
above results, we have for this quantity that:
\begin{equation}
d_{eff}=\left\{
\begin{array}
[c]{c}%
\sim n\text{ for maximally mixed symmetric state}\\
\sim\sqrt{n}\text{ for pure symmetric state}\\
\sim n^{1/4}\text{ for critical XXZ\ model ground state}\\
\sim n^{c/4}\text{ for a critical state with central charge }c\text{ }%
\end{array}
\right. \nonumber
\end{equation}

From the expression of $Tr(\rho_{(n)}^{\alpha})$ in \ref{tra_symm} the R\'enyi
and Tsallis entropies for fully symmetric states follow as
\begin{align}
\label{renyi-heis}S_{R}(\alpha, n)  &  =\frac{1}{2}\log(2\pi n pq)-\frac
{\log\alpha}{2(1-\alpha)},\\
S_{T}(\alpha,n)  &  =\frac{1}{\sqrt{\alpha}} \frac{(2\pi n p q)^{\frac
{1-\alpha}{2}}-1}{1-\alpha}. \label{tsallis-heis}%
\end{align}
In Figs. \ref{Fig_renyi}, \ref{Fig_tsallis}, we compare the above analytical
expressions for the R\'enyi and Tsallis entropies with exact calculations
using the RDM eigenvalues in Eqs (\ref{eigenvalue_basis_coefficients})
(\ref{alpha_Z_general}), from which we see that a very good agreement is
found. Also notice that in the limit $\alpha=1$ both entropies reduce to the
entanglement entropy (\ref{VNE(T)}) at $T=0$: $S_{R}(1,n)=S_{T}(1,n)=\frac
{1}{2}\log2\pi epqn$.

In general, for arbitrary permutational symmetries and for finite
temperatures, one must recourse to direct calculations using the general
expression (\ref{eigenvalue_basis_coefficients}) for the RDM eigenvalues,
since it is not easy in these cases to give simple analytical expressions of
$\alpha$. The study of the analytical properties of the RDM moments represents
an interesting problem which deserves further investigations.

\section{Conclusions}

To summarize, we have provided explicit analytical expression of the reduced
density matrix of a subsystem of arbitrary size $n$ of a permutational
invariant quantum many body system of arbitrary size $L$ and characterized by
a state of arbitrary permutational symmetry. We have shown, on the specific
example of the spin $1/2$ Heisenberg model, that the RDM acquires a block
diagonal form with respect to the quantum number $k$ fixing the polarization
in the subsystem conservation of $S_{z}$) and with respect to the irreducible
representations of the $\mathbf{S_{n}}$ group. Analytical expression for the
RDM elements and for the RDM spectrum are derived for states of arbitrary
permutational symmetry and for arbitrary fillings. These results are provided
by Eqs. (\ref{g_z}), (\ref{Gz_elements_thermodynamic_limit}) and
(\ref{alpha_Z_general}) presented above. Entanglement properties have been
discussed both in terms of the VNE and of the Renyi and Tsallis entropies. In
particular, the temperature dependence and the scaling of the VNE across a
finite temperature phase transition have been considered and the RDM moments
and the R\'enyi and Tsallis entropies have been calculated for symmetric
ground states of the Heisenberg chain and for maximally mixed states. These
results being based only on the permutational invariance and on the
conservation of $S_{z}$ (number of particles for non spin systems) are
expected to apply also to other quantum many-body systems with the same
symmetry properties.

\vskip1 cm \textbf{Acknowledgments}
This paper is written  in honor of the 60th birthday of Professor Vladimir Korepin. V.P. thanks V. Korepin and V. Vedral for stimulating discussions, the Center for Quantum
Technologies, Singapore for hospitality, and  the University of Salerno for a
research grant (Assegno di Ricerca no. 1508). M. S. acknowledges support from the
Ministero dell' Istruzione, dell' Universit\`{a} e della Ricerca (MIUR)
through a \textit{Programma di Ricerca Scientifica di Rilevante Interesse
Nazionale} (PRIN)-2008 initiative. 

\appendix

\section{RDM elements for symmetric states}

\label{app::RDM elements for symmetric states}

In this appendix we provide a proof of Eq. (\ref{g_z}) which is valid for
fully symmetric states (case $r=0$) such as, for example, the ground state of
the ferromagnetic Heisenberg chain. For $r=0$, the corresponding YT is
nondegenerate and the state of the full system is pure: $\rho=|\Psi
_{L,N}\rangle\langle\Psi_{L,N}|$ with $|\Psi_{L,N}\rangle$ the symmetric
state
\begin{equation}
|\Psi_{L,N}\rangle=\binom{L}{N}^{-1/2}\sum_{P}|\underbrace{\uparrow
\uparrow...\uparrow}_{N}\underbrace{\downarrow\downarrow...\downarrow\rangle
}_{L-N} \label{GS}%
\end{equation}
where the sum is over all possible permutations. Since all $L$ sites are
equivalent due to permutational invariance, any choice of $n$ sites
$i_{1},i_{2},...,i_{n}$ within $L$ sites gives the same RDM, which we denote
by $\rho_{(n)}^{L,N,0}=Tr_{L-n}\rho$. It has been shown in
\cite{Entanglement_Heisenberg} that $\rho_{(n)}^{L,N,0}$ takes form
\begin{equation}
\rho_{(n)}^{L,N,0}=\sum_{k=0}^{n}\frac{\binom{L-n}{N-k}}{\binom{L}{N}}
|\Psi_{n,k}\rangle\langle\Psi_{n,k}|.\nonumber
\end{equation}
In the natural basis the matrix elements of RDM are given by the above
discussed values of observables. Using (\ref{GS}), one explicitly computes all
RDM elements as
\begin{equation}
\left(  \rho_{(n)}^{L,N,0}\right)  _{Q}^{P}=\left(  \rho_{(n)}^{L,N,0}\right)
_{i^{\prime}j^{\prime}...m^{\prime}}^{ij...m}=\delta_{i^{\prime}+j^{\prime
}...+m^{\prime}}^{i+j+..+.m}\frac{\binom{L-n}{N-w}}{\binom{L}{N}}
\label{RDM_r=0}%
\end{equation}
with $w=i+j+...+m$ (the sets $ij...m$ and $i^{\prime}j^{\prime}...m^{\prime}$
are binary representation of numbers $P-1,Q-1$). In this case we have that all
elements of a block $B_{k}$ are equal (this is not true for $r\geq0$). We also
see that the the elements in Eq. (\ref{RDM_r=0}) are the same as those
obtained from Eq. (\ref{g_z}) for $r=0$. Note that in the thermodynamic limit
$\eta=1$, and $\binom{L-n}{N-k}/\binom{L}{N}\rightarrow p^{n-k}(1-p)^{k}$ in
agreement with Eq. (\ref{Gz_elements_thermodynamic_limit}).

\section{RDM elements in the thermodynamic limit}

\label{app::RDM elements in the thermodynamic limit}

To calculate the RDM, we shall use the representation
(\ref{sigma_as_SumOfPermutaitons}) for the density matrix of the whole system
$\sigma$, rewritten in the form
\begin{align}
\rho_{(n)}  &  =Tr_{L-n}\left\{  \frac{1}{L!}\sum_{P}|\Psi_{L,N,r}%
\rangle\langle\Psi_{L,N,r}|\right\} \nonumber\\
&  =Tr_{L-n}\left\{  \frac{1}{n!}\frac{1}{(L-n)!}\frac{1}{\binom{L}{n}}%
\sum_{P_{(n)}}\sum_{P_{(L-n)}}\right. \nonumber\\
&  \left.  \sum_{i_{1}\neq i_{2}\neq...\neq i_{n}}|\Psi_{L,N,r}\rangle
\langle\Psi_{L,N,r}|\right\}  . \label{PermutationSplittingInSteps}%
\end{align}
Note that the $L!$ permutations can be done in three steps: first, choose at
random $n$ sites $i_{1}\neq i_{2}\neq...\neq i_{n}$ among the $L$ sites. There
are $\binom{L}{n}$ such choices. Then, permute the chosen $n$ sites, the total
number of such permutations being $n!$. Finally, permute the remaining $L-n$
sites, the total number of such permutations being $(L-n)!$. The latter step
(c) under the trace operation is irrelevant because these degrees of freedom
will be traced out. The operation permuting $n$ sites commutes with the trace
operation since $Tr_{L-n}$ does not touch the respective subset of $n$ sites.
Consequently, (\ref{PermutationSplittingInSteps}) can be rewritten as
\begin{equation}
\rho_{(n)}=\frac{1}{n!}\sum_{P_{(n)}}Tr_{L-n}\frac{1}{\binom{L}{n}}\sum
_{i_{1}\neq i_{2}\neq...\neq i_{n}}|\Psi_{L,N,r}\rangle\langle\Psi_{L,N,r}|\,.
\label{RDM_operational}%
\end{equation}
We recall here that a filled $YT$ of type $\{L-r,r\}_{(N)}$ contains a mixed
symmetry part with $2r$ sites and $r$ spin up , and a fully symmetric part
with $L-2r$ sites and $N-r$ spin up (in the following we adopt an equivalent
terminology which refers to spins up as particles and to a spins down as
holes). This implies that the corresponding wave function $|\Psi
_{L,N,r}\rangle$ factorizes into symmetric and antisymmetric parts as
\begin{equation}
\Psi=|\phi_{12}\rangle\otimes|\phi_{34}\rangle\otimes...\otimes|\phi
_{2r-1,2r}\rangle\otimes|\Psi_{L-2r,N-r}\rangle_{2r+1,2r+2,..L}
\label{YoungTableau_example}%
\end{equation}
with the antisymmetric part consisting of the first $r$ factors of the type
\begin{equation}
|\phi_{12}\rangle=\frac{1}{\sqrt{2}}(|10\rangle_{12}-|01\rangle_{12})
\label{asym_wavefunction}%
\end{equation}
and with the symmetric part, $|\Psi_{L-2r,N-r}\rangle$, given by (\ref{GS}). A
general property of factorized states implies that if the global wave function
is factorized, $|\Phi\rangle=|\psi\rangle_{I}|\phi\rangle_{II}$ and out of $n$
sites of the subsystem, $n_{1}$ sites belong to subset $I$, and the remaining
$n_{2}=n-n_{1}$ sites belong to the subset $II$, then the reduced density
matrix factorizes as well:
\begin{equation}
\rho_{(n)}=\rho_{(n_{1})}^{I}\otimes\rho_{(n_{2})}^{II}.
\label{density_matrix_product}%
\end{equation}

To do the averaging, we note that among total number of choices $\binom{L}{n}
$ there are (a) $\binom{L-2r}{n}$ possibilities to choose $n$ sites inside the
symmetric part of the tableau, containing $N-r$ particles, (b) $2r\binom
{L-2r}{n-1}$ possibilities to choose $n-1$ sites inside the symmetric part of
the tableau and one site in the antisymmetric part (c) $\binom{2r}{2}%
\binom{L-2r}{n-2}$ possibilities to choose $n-2$ sites inside the symmetric
part of the tableau and two sites in the antisymmetric part and so on. The
contributions given by (a) and (b) to the right hand side of
(\ref{RDM_operational}) for $\rho_{(n)}^{L,N,r}$ are given, according to
(\ref{density_matrix_product}), by
\begin{equation}
\left\langle \binom{F}{n}\rho_{(n)}^{F,M,0}+2r\binom{F}{n-1}\rho
_{(n-1)}^{F,M,0}\otimes\rho_{\frac{1}{2}}\right\rangle ,
\end{equation}
with $F=L-2r$, $M=N-r$ and with $\rho_{\frac{1}{2}}=I/2$ the density matrix
corresponding to a single site in the antisymmetric part of the tableau.
Brackets $\langle.\rangle=\frac{1}{n!}\sum_{P}$ denote the average with
respect to permutations of $n$ elements.The contribution \ due to (c) to
(\ref{RDM_operational}) splits into two parts since the $\binom{2r}{2}$
possibilities to choose two sites in the antisymmetric part of the tableau
consist of $4 \binom{r}{2}$ choices with two sites into different columns and
the remaining $r$ choices with both sites belonging to a same column. For the
former choice, the corresponding density matrix is $\rho_{(n-2)}%
^{F,M,0}\otimes\rho_{\frac{1}{2}}\otimes\rho_{\frac{1}{2}}$, while for the
latter case is given by $\rho_{(n-2)}^{F,M,0}\otimes\rho_{asymm}$, with
\begin{equation}
\rho_{asymm}=|\frac{1}{\sqrt{2}}(10-01)\rangle\langle\frac{1}{\sqrt{2}%
}(10-01)|=\frac{1}{2}%
\begin{pmatrix}
0 & 0 & 0 & 0\\
0 & 1 & -1 & 0\\
0 & -1 & 1 & 0\\
0 & 0 & 0 & 0
\end{pmatrix}
. \label{rho_asymm}%
\end{equation}
Proceeding in the same manner for arbitrary partitions of $Z$ sites in the
antisymmetric part of the tableau and $n-Z$ sites in the symmetric part, we
get
\begin{align}
&  \rho_{(n)}^{L,N,r} \binom{L}{n} = \left\langle \sum_{Z=0}^{\min(2r,n)}
\binom{F}{n-Z} \rho_{(n-Z)}^{F,M,0} \sum_{i=0}^{[Z/2]}\binom{r}{i} \right.
\nonumber\\
&  \left.  \left(  {\prod\limits_{1}^{i}}\otimes\rho_{asymm}\right)  2^{Z-2i}
\binom{r-i}{Z-2i}\left(  {\displaystyle\prod\limits_{1}^{Z-2i}}\otimes
\rho_{\frac{1}{2}}\right)  \right\rangle .
\label{thermal_density_matrix_decomposition1}%
\end{align}
From this the general scheme for the decomposition of the general RDM becomes
evident. In the above formula, the products $\prod\limits_{i}^{Q}$ with $Q<i$
are discarded. The matrix elements $\rho_{(k)}^{F,M,0}$ are given by
(\ref{RDM_r=0}).

For simplicity of presentation, we prove Eq.
(\ref{Gz_elements_thermodynamic_limit}) for the case $Z=k$ and then outline
the proof for arbitrary $Z$.

In the thermodynamic limit one can neglect the difference between factors like
$4\binom{r}{2}$ and $\binom{2r}{2}$ in Eq.
(\ref{thermal_density_matrix_decomposition1}). The latter then can be then
rewritten in a simpler form as
\begin{align}
\binom{L}{n}\rho_{(n)}^{L,N,r}  &  =\left\langle \binom{F}{n}\rho
_{(n)}^{F,M,0}+\binom{2r}{1}\binom{F}{n-1}\rho_{(n-1)}^{F,M,0}\otimes
\rho_{\frac{1}{2}}\right.
\label{thermal_density_matrix_thermodynamic_limit_decomposition}\\
&  \left.  +\binom{2r}{2}\binom{F}{n-2}\rho_{(n-2)}^{F,M,0}\otimes\rho
_{\frac{1}{2}}\otimes\rho_{\frac{1}{2}}+...\right\rangle .\nonumber
\end{align}
Note that one can omit all terms in
(\ref{thermal_density_matrix_decomposition1}) containing $\rho_{asymm}$ since
the respective coefficients correspond to probabilities of finding two
adjacent sites in the asymmetric part of the YT (proportional to $r$), which
vanish in the thermodynamic limit, respect to the total number of choices
which is of order of $r^{2}$. A sub-block $G_{Z}$ of a block $k$ consists of
all elements of the matrix $\rho_{(n)}$ having $Z$ pairs of $e_{1}^{0}%
,e_{1}^{0}$ in its tensor representation, like e.g. $\left(  e_{1}^{0}\otimes
e_{0}^{1}\right)  ^{\otimes_{Z}}\otimes e_{i_{1}}^{i_{1}}\otimes e_{i_{2}%
}^{i_{2}}\otimes...\otimes e_{i_{n-2Z}}^{i_{n-2Z}}$, such that $Z+i_{1}%
+i_{2}+...+i_{n-2Z}=k$. The total number of elements $g_{Z}\subset G_{Z}$ in
$\rho_{(n)}^{L,N,r}$ is equal to the number of distributions of $Z$ objects
$e_{1}^{0}$, $Z$ objects $e_{0}^{1}$, and $(k-Z)$ objects $e_{1}^{1}$ on $n$
places, given by
\begin{equation}
\deg G_{Z}=\frac{n!}{Z!Z!(k-Z)!(n-k-Z)!}%
\end{equation}
(this is another way of writing (\ref{deg_Gz(k)})). Each term $W$ in the sum
(\ref{thermal_density_matrix_thermodynamic_limit_decomposition}) after
averaging will acquire the factor
\begin{equation}
\Gamma(W)=\frac{\deg G_{Z}(W)}{\deg G_{Z}} \label{averaging_factor}%
\end{equation}
where $\deg G_{Z}(W)$ is a total number of $g_{Z}$ elements in the term $W$,
provided all of them are equal. For instance, $\deg G_{Z}(\rho_{(n)}%
^{F,M,0})=\deg G_{Z}$, \ $\deg G_{Z}(\rho_{(n-m)}^{F,M,0}\otimes\left(
\rho_{\frac{1}{2}}\right)  ^{\otimes_{m}})=\binom{2Z}{Z}\binom{n-m}{2Z}%
\sum_{m_{1}=0}^{m}\binom{m}{m_{1}}\binom{n-2Z-m}{k-Z-m_{1}}$( the last formula
is only true for $k=Z$, otherwise elements constituting $G_{Z}(W)$ are not all
equal). Restricting to the case $k=Z$ and denoting $W_{m}=\rho_{(n-m)}%
^{F,M,0}\otimes\left(  \rho_{\frac{1}{2}}\right)  ^{\otimes_{m}}$, we have
\begin{equation}
\Gamma(W_{m})=\Gamma_{m}=\frac{\binom{n-m}{2Z}}{\binom{n}{2Z}}.
\label{averaging_factor_m}%
\end{equation}
It is worth to note that the element $g_{Z}\subset G_{Z}$ is simply given by
\begin{align}
\binom{L}{n}g_{Z}=\Gamma_{0}\binom{F}{n}g_{0}^{(n,k)}+\Gamma_{1}\binom{2r}%
{1}\binom{F}{n-1}\frac{g_{0}^{(n-1,k)}}{2} +\nonumber\\
\Gamma_{2}\binom{2r}{2}\binom{F}{n-2}\frac{g_{0}^{(n-2,k)}}{2^{2}}+ ...
\qquad\qquad\qquad\qquad\label{c_element_thermodynamic_limit}%
\end{align}
with $q=1-p$ and $g_{0}^{(n,k)}=\frac{\binom{F-n}{M-k}}{\binom{F}{M}}%
\approx\left(  \frac{p-\mu}{1-2\mu}\right)  ^{^{n-k}}\left(  \frac{q-\mu
}{1-2\mu}\right)  ^{k}$ is the element of a $\rho_{(n)}^{F,M,0}$ corresponding
to a block with $k$ particles (the factors $\Gamma_{m}$ are due to the
averaging while the factors $\frac{1}{2^{m}}$ come from $\left(  \rho
_{\frac{1}{2} }\right)  ^{\otimes_{m}}$). Restricting to the case $k=Z$, and
taking into account
\begin{equation}
\frac{\binom{F}{n-m}}{\binom{L}{n}} \approx\frac{n!}{(n-m)!}\frac{(1-2\mu
)^{n}}{F^{m}}, \;\;\;\;\;\;\; \binom{2r}{m} \approx\frac{(2\mu)^{m}}{m!}%
L^{m}\;,
\end{equation}
so that
\[
\frac{\binom{F}{n-m}}{\binom{L}{n}}\binom{2r}{m}\frac{1}{2^{m}}\approx
\binom{n}{m}\mu^{m}(1-2\mu)^{n-m},
\]
we finally obtain, using (\ref{c_element_thermodynamic_limit}), that
\begin{align}
g_{Z}  &  = \sum_{m=0}^{n-2Z}\mu^{m}\binom{n-2Z}{m}(p-\mu)^{n-m-Z}(q-\mu
)^{Z}\nonumber\\
&  = (p-\mu)^{n-Z}(q-\mu)^{Z}\sum_{m=0}^{n-2Z}\frac{\mu^{m}}{(p-\mu)^{m}}
\binom{n-2Z}{m}\nonumber\\
&  = (p-\mu)^{n-Z}(q-\mu)^{Z}\left(  \frac{p}{p-\mu}\right)  ^{n-2Z}%
\label{g_Z_thermodynamic_k=Z}\\
&  = p^{n-Z}q^{Z}\left(  \frac{(p-\mu)(q-\mu)}{pq}\right)  ^{Z}\nonumber\\
&  = p^{n-Z}q^{Z}\eta^{Z} = \eta^{Z}g_{0} \,,\nonumber
\end{align}
with $g_{0}$ the diagonal element in the same block $k=Z$. In the last
calculation we used the relation $\frac{\binom{n}{m}\binom{n-m}{2Z}}{\binom
{n}{n-Z}}=\binom{n-2Z}{m}$. This proves formula
(\ref{Gz_elements_thermodynamic_limit}) for the particular case $k=Z$ and
arbitrary $n$.

For arbitrary $k,Z,$ one proceeds in similar manner as for the case $k=Z$
case. Since the respective calculations are tedious and not particularly
illuminating, we omit them and give only the final result:
\begin{align}
g_{Z}  &  = \sum_{m=0}^{n-2Z}\mu^{m}\sum_{i=\max(0,Z+k-n+m)}^{\min(m,k-Z)}(p-
\mu)^{n-m-k+i}\nonumber\\
&  (q-\mu)^{k-i}\binom{k-Z}{i}\binom{n-k-Z}{m-i},
\end{align}
which, after some algebraic manipulation, can be rewritten in the form
\begin{align}
g_{Z}  &  = (p-\mu)^{n-k}(q-\mu)^{k}\sum_{j=0}^{n-k-Z}\left(  \frac{\mu}{
p-\mu}\right)  ^{j}\binom{n-k-Z}{j}\nonumber\\
&  \times\sum_{i=0}^{k-Z}\left(  \frac{\mu}{q-\mu}\right)  ^{i}\binom{k-Z}{i}
= \eta^{Z}p^{n-k}q^{k}. \label{g_Z_thermodynamic}%
\end{align}
This concludes the proof of Eq. (\ref{g_z}) in the thermodynamic limit
$L\rightarrow\infty$.

\end{document}